\documentclass{aa}
\usepackage{graphicx}
\begin{document}

\newcommand{\inthms}[3]{$#1^{\rm h}#2^{\rm m}#3^{\rm s}$}
\newcommand{\dechms}[4]{$#1^{\rm h}#2^{\rm m}#3\mbox{$^{\rm
s}\mskip-7.6mu.\,$}#4$}
\newcommand{\intdms}[3]{$#1^{\circ}#2'#3''$}
\newcommand{\decdms}[4]{$#1^{\circ}#2'#3\mbox{$''\mskip-7.6mu.\,$}#4$}
\newcommand{\tmb}{\mbox{T$_{\rm mb}$}}
\newcommand{\OI}{\mbox{O\,{\sc i}}}
\newcommand{\dtco}{D$_{2}$CO}
\newcommand{\hdco}{HDCO}
\newcommand{\htco}{H$_{2}$CO}
\newcommand{\httco}{H$_{2}^{13}$CO}
\newcommand{\kmps}{km s$^{-1}$}

\hyphenation{Fe-bru-ary Gra-na-da mo-le-cu-le mo-le-cu-les}

%\thesaurus{08.02.3:08.03.4:08.06.2:08.16.5:02.01.2}

\title{Observational constraints on the afterglow of GRB 020531
}

\author{
        A. Klotz\inst{1}
        \and M. Bo\"er\inst{1}
        \and J.L. Atteia\inst{2}
} \institute{ Centre d'Etude Spatiale des Rayonnements,
Observatoire Midi-Pyr\'en\'ees (CNRS-UPS), BP 4346, F--31028 -
Toulouse Cedex 04, France \and Laboratoire d'Astrophysique,
Observatoire Midi-Pyr\'en\'ees (CNRS-UPS), 14 Avenue E. Belin,
F--31400 - Toulouse, France }

\offprints{A. Klotz, \email{klotz@cesr.fr}}

\date{Received November 19th, 2002 /Accepted }

\titlerunning{Observational constraints on the afterglow of GRB 020531}
\authorrunning{Klotz {\it et al.}}

\abstract{ We present the data acquired by the TAROT automated
observatory on the afterglow of GRB 020531.
Up to now, no convincing
afterglow emission has been reported for this short/hard GRB at
any wavelength, including X-ray and optical. The combination of
our early limits, with other published data allows us to
put severe constraints on the afterglow magnitude and light curve.
The limiting magnitude is
18.5 in R band, 88 minutes after the GRB, and the decay slope power
law index could be larger than 2.2.}
 \maketitle

%%%%%%%%%%%%%%%%%%%%%%%%%%%%%%% INTRODUCTION %%%%%%%%%%%%%%%%%%%%%%%%%
\section{Introduction}
Since their first detection by van Paradijs
{\it et al.} \cite{Vanpara97}, gamma-ray burst (GRB)
optical afterglows have been
detected in about 40\% of the sources displaying an X-ray
afterglow. The fireball model (Rees and M\'esz\'aros
\cite{Rees92}, M\'esz\'aros and Rees \cite{Mesz97}, Panaitescu
{\it et al.} \cite{Pana98}) has been established as a
standard tool to
interpret these observations. In this framework the afterglow
emission is described as synchrotron and inverse Compton emission
of high energy electrons accelerated during the shock of an
ultra-relativistic shell with the external medium, while the
prompt emission is due to the internal shocks produced by shells
of different Lorentz factors
within the relativistic blast wave (see Piran
\cite{Piran99} for a review). Both the prompt radiation and early
afterglow phases provide critical information to establish the
physical processes at work during the burst itself, as well as the
physical conditions of the surrounding environment (Kumar and
Panaitescu \cite{Kum00}, Kumar and Piran \cite{Kum2000b}). There
is a general consensus that the fireball plasma is constituted by
$e^-e^+$ pairs and $\gamma$-ray photons, however the ultimate energy
reservoir and the detailed radiation
mechanisms are still a challenge
to theoretical models.

The situation of 60\% of the GRB afterglows
which are not observed at optical
wavelengths (called {\it dark GRBs}) is not clear. As it has
been shown in Bo\"er and Gendre (\cite{Boer00}), the optical flux
is not correlated with the intensity of the X-ray afterglow, nor
with the distance. Generally speaking the absence of an optical
transient associated with a GRB can be attributed to
four, non exclusive, reasons, namely 1) the distance of the
source, though this is obviously not the general case, 2) the
absorption of the visible light by a dense medium (
{\it I.E.} dust),
3) the rapid
decay of the optical afterglow, and 4) the intrinsic faintness of
the source at long wavelengths ({\it i.e.} optical, NIR...). However,
a few reports of near IR and optical non-detection of GRB
afterglows show, that hypothesis 2 is not the main reason
(see {\it e.g.} GRB 010214, Piro \cite{Piro2001} and
subsequent GCN circular available at the URL
http://gcn.gsfc.nasa.gov/gcn/other/010214.gcn3). In the absence of
rapid simultaneous X-ray and optical measurements, hypotheses 3 and
4 are difficult to evaluate.

It should be noted that for the sub-class of GRBs
that exhibit a short duration and a hard spectrum,
usually called {\it short/hard GRB}
(Dezalay {\it et al.} \cite{dezalay96}, Kouveliotou {\it et al.}
\cite{kou93}),
no optical counterpart has been detected yet
(Hurley {\it et al.} \cite{hurley2002a},
Gorosabel {\it et al.} \cite{Goro02}).
This is largely due to the scarcity of the observations. If this
appears a "general" law, it can be the indication of
a different geometry (as viewed from the observer)
or of another mechanism
for the emission of the afterglow
({\it e.g.} Shanthi {\it et al.} \cite{Shanthi99}).
Hence, it is important to get
rapid and deep measures (or upper limits) on the afterglow
emission for GRB sources of all classes, and particulary
for the short GRBs.

In this letter we report on the early observations of GRB 020531
performed with the automatic TAROT observatory (Bo\"er {\it et
al.} \cite{Boer99}). Our data, combined with the data from other
telescopes strongly constrain both the magnitude and the decay
slope index of the optical counterpart, if any.

%%%%%%%%%%%%%%%%%%%%%%%%%%%%%%% OBSERVATIONS %%%%%%%%%%%%%%%%%%%%%%%%%
\section{Observations}
\label{obsdata}

\subsection{Detection and follow up of the burst}
\label{grb020531}

The High Energy Transient Explorer satellite
(HETE, Ricker {\it et al.} \cite{ricker2000}) detected GRB
020531 with the FREGATE and WXM instruments on
May 31, 2002 at 0h26min18.73 UTC (Ricker {\it et al.}
\cite{ricker2002}). This event is a short/hard GRB:
t$_{90}$=0.94s, t$_{50}$=0.45s, and fluence is 8$\cdot$10$^{-7}$
erg cm$^{-2}$ in the FREGATE
50-300 keV band. The absolute localization was not performed by the
flight software and the preliminary coordinates were computed by a
ground analysis. The GRB Coordinates Network (GCN - Barthelmy
\cite{Bar97}) broadcasted the position at 1h54min22s UT. Twenty-five GCN
circulars (GCNC) were published on this event between May 31 and
July 25, 2002. In the first very early reports, it appears that no
unambiguous optical counterpart was recorded. Five days after the
GRB, only four faint sources were detected by the Chandra
satellite ACIS-I array (Butler {\it et al.} \cite{butler2002}) in the
final error box given by the Inter Planetary Network (IPN) published
on the July 10$^{th}$ 2002 (Hurley {\it et al.}
\cite{hurley2002b}). The connection of one of these X-ray sources
with the gamma-ray transient remains to be confirmed.
Complementary informations about the GRB localization
can be found in Lamb {\it et al.} \cite{lamb2002}.

%%%%%%%%%%%%%%%%%%% FIG1 %%%%%%%%%%%%%%%%%%%
\begin{figure}[t]
\includegraphics[width=\columnwidth]{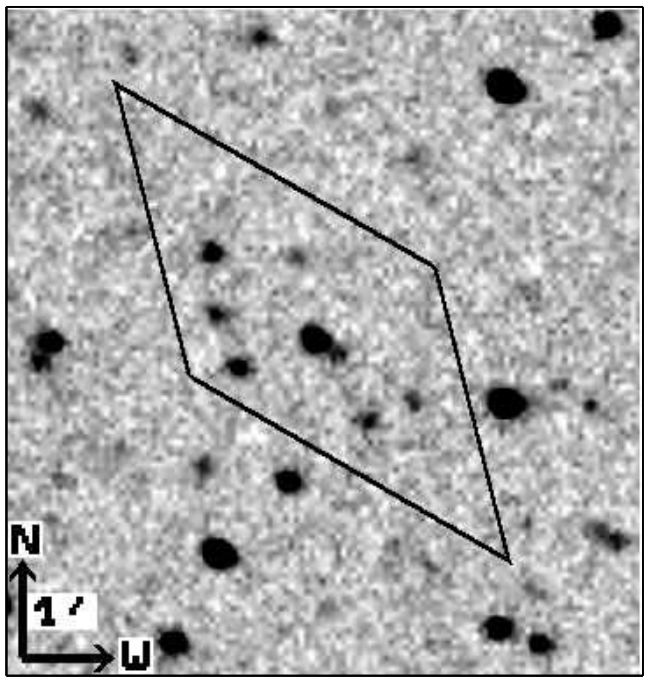}
\caption{ A sub-image of the TAROT composite image (11 frame of
duration 30 seconds each). The parallelogram is the last IPN error
box (from GCNC 1461, Hurley {\it et al.} \cite{hurley2002b}). }
\label{TAROTimage}
\end{figure}
%

%%%%%%%%%%%%%%%%%%% FIG2 %%%%%%%%%%%%%%%%%%%
\begin{figure}[t]
\includegraphics[width=\columnwidth]{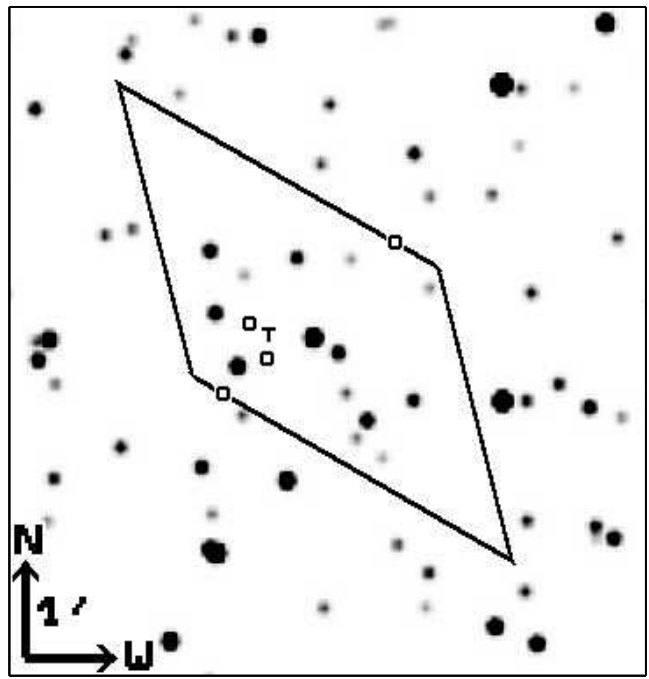}
\caption{ A synthetic Cousin R image of the same field than  the
TAROT image in figure 1. Magnitude values are taken from the
BVRcIc all-sky photometry, posted by Henden \cite{henden2002} in
GCNC 1422. Only stars brighter than Rc=18.5 are displayed, i.e. up
to the limiting magnitude of the TAROT image. Circles indicate the
positions of the Chandra observatory X-ray sources (from GCNC
1415, Butler {\it et al.} \cite{butler2002}). The T symbol is the
location of the Tarot-C source (from GCNC 1420, Klotz {\it et
al.} \cite{klotz2002}). }
\label{SyntheIma}
\end{figure}
%

%\subsection{This study}
\label{thisstudy} Up to now, none of the suggested optical
counterparts of GRB 020531 has been confirmed. In this study we
present the data acquired with the TAROT observatory. Our limits
are compared with the limiting magnitudes obtained by other
observers at different times after the GRB. Given that our data
were obtained only 88 minutes after the burst itself, we can infer
strong limits both on the optical counterpart magnitude and decay
slope.

%%%%%%%%%%%%%%%%%%%%%%%%%%% TAROT %%%%%%%%%%%%%%%%%%%%%%%%%
\subsection{Tarot observations}
\label{tarotobs} TAROT is a fully autonomous 25 cm aperture
telescope installed at the Calern observatory (Observatoire de la
Cote d'Azur - France). Its $2\degr$ field of view
ensures the total coverage of HETE error boxes.
This telescope is
devoted to very early observations of GRB optical counterparts.
A technical description of TAROT can be read in Bo\"er {\it et
al.} \cite{Boer99} and in Bringer {\it et al.} \cite{bringer2001}. The
CCD camera is based on a THX7899 Thomson chip. It is placed at the
newtonian focus. The focal length is 0.81 meter and the pixel size
is 14 microns. The spatial sampling is 3.5 arcsec/pixel. The readout
noise is 13 electrons rms and the actual gain is 3.6
photo-electrons/adu. The main feature of this camera is its very
short readout time: 2 seconds to read the entire 2048x2048 matrix
with no binning.

The first image was taken by TAROT less than 6 seconds after the
position of GRB 020531 was provided by the GCN. A series of 11
unfiltered images of 30 seconds was then taken. An automatic
preprocessing software gave scientific images in the following
minutes. We compared them to the Digital Sky Survey (DSS) images.
We concluded quickly that no bright new source was present. The
limiting magnitude of the individual images, in the Cousin R band,
is about 16.7.
\\
Then we coadded the 11 images to improve the signal to noise
ratio (see figure 1). A limiting magnitude of 18.5 (compared
to the R cousin
band) is reached. This limiting magnitude is estimated from
comparison with a set of synthetic images computed from the BVRcIc
USNOFS all-sky photometry of field (Henden \cite{henden2002}). On
figure 2, only stars brighter than Rc=18.5 are plotted.
\\
Three TAROT sources, afterglow candidates, were published in
the GCN circulars : sources A and
B (Bo\"er {\it et al.} \cite{boer2002}) and C (Klotz {\it et al.}
\cite{klotz2002}).

%%%%%%%%%%%%%%%%%%%%%%%%%%% TABLE  %%%%%%%%%%%%%%%%%%%%%%%%%%%%%
\begin{table}[htb]
\caption{ Log of the published values of the limiting magnitudes,
presented in the chronological order. The first column is the date
from GRB (in days). The second is the limiting R magnitude of
the image. The third is the GCN
circular index of the publication. }
\begin{center}
\begin{tabular}{c c c c}
Date & R lim      & GCNC & Instrument \cr
\noalign{\smallskip} \hline \noalign{\smallskip} 0.0654&18.5&1408
& TAROT (D=0.25 m) \cr 0.0997   & 17.7  & 1406 & D. West (D=0.20
m)  \cr 0.1512   & 17.5  & 1404 & Super-LOTIS (D=0.60 m) \cr
0.1831   & 18    & 1400 & NEAT (D=1.2 m)\cr
0.1859   & 18 & 1401 &
SDSS (D=0.5 m)  \cr 0.1873   & 20.5  & 1405 & KAIT (D=0.8 m) \cr
0.3790   & 18    & 1401 & SDSS (D=0.5 m)  \cr 0.9017   & 24.7 &
1433 & INT (D=2.5 m)  \cr 1.1417   & 23.6  & 1434 & Baade (D=6.5
m)  \cr 1.2352   & 20.5  & 1405 & KAIT (D=0.8 m)  \cr 2.9717   &
25.2  & 1433 & INT (D=2.5 m)  \cr 5.4317   & 25.5  & 1434 & Subaru
(D=8.2 m)  \cr 10.1117  & 24.0  & 1434 & Baade (D=6.5 m)  \cr
\noalign{\smallskip} \hline
\end{tabular}
\label{logobstable}
\end{center}
\end{table}

Source A, RA=15h14min51s DEC=-19$\degr$25'06" (J2000.0), R=17.4,
cannot be the asteroid number 2 mentioned by Li {\it et al.}
\cite{li2002} in the GCNC 1405, as it was supposed by Bo\"er {\it et
al.} \cite{boer2002} in the GCNC 1408. The reason is that it lies
in the opposite side of the apparent motion published by Li {\it et al.}
\cite{li2002}. Source B, RA=15h14min57s DEC=-19$\degr$28'12"
(J2000.0), R=17.1, is a known star visible in DSS and various
other images. Anyway, A and B sources lie outside the IPN error
box.
\\
Source C, RA=15h15min12s DEC=-19$\degr$24'24" (J2000.0), R$\geq$
18.5, is considered as the best TAROT image candidate in the IPN
error box. We reprocessed the raw images using the calibration
frames taken both before and after the night of May 30-31, 2002,
and we obtained a fainter source on the new refined co-added
images. This meant that source C could be a group of "hot pixel"
badly corrected by the automatic preprocessing which uses only the
calibration frame taken during the preceding day, to produce
synthetic calibration data.

%%%%%%%%%%%%%%%%%%% FIG1 %%%%%%%%%%%%%%%%%%%
\begin{figure}[htb]
\includegraphics[width=\columnwidth]{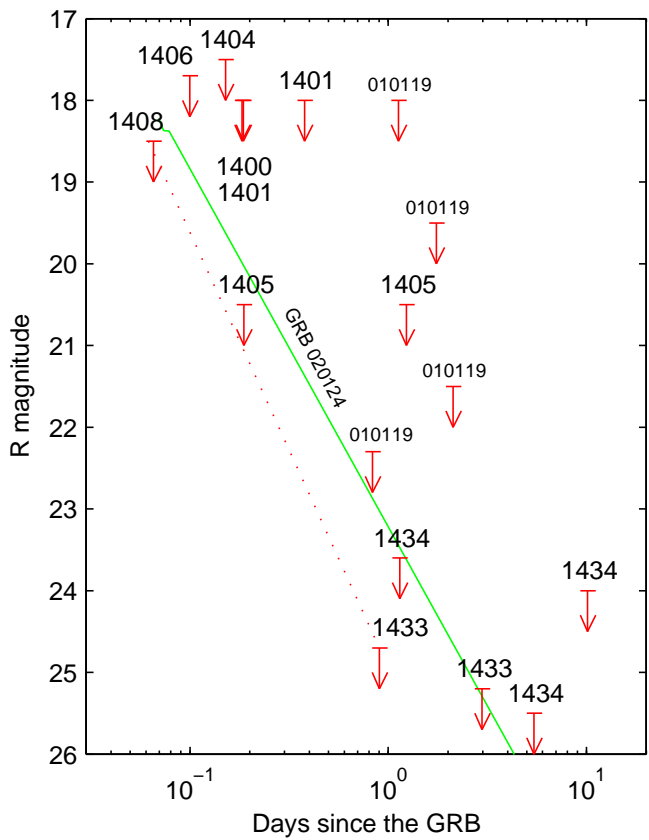}
\caption{Reported lower limits of the magnitude (from table 1) of
the optical afterglow of GRB 020531 (arrows, with the GCN circular
number). For comparison, we added some data (labeled 010119)
of the upper limits
of the short/hard GRB 010119 (Gorosabel {\it et al.}
\cite{Goro02}) and we plotted, as a solid line, the light curve
of the dim afterglow of the long burst GRB 020124
(Berger {\it et al.} \cite{Berger02}).
The dashed line represent the upper limit for the
brightness of the afterglow, assuming a constant
decay slope.
} \label{curves}
\end{figure}

Other fuzy patches are also seen in the image of
TAROT presented in figure 1. All of these patches
can be related to known stars fainter than Rc=18.5.
However, as the TAROT image is unfiltered, it is not
surprising to find these stars (color effects).

%%%%%%%%%%%%%%%%%%%%%%%%%%% OTHERS %%%%%%%%%%%%%%%%%%%%%%%%%
\subsection{Other Observations}
\label{otherobs} The data reported in various GCN circulars are
summarized on  table 1. The first column is the delay, in fraction
of day, between the burst and the beginning of the observation,
the second column gives the limiting magnitude, the third
column indicates the GCN circular in which the data was reported,
and the last one the instrument used as well as its aperture. For
early observations ($<$ 1 day after GRB), only small aperture
telescopes (i.e. $<$ 2 meters) scanned the field. During this
delay, the better limiting magnitude is 20.5 from the Katzman
Automatic Imaging Telescope (KAIT, Li {\it et al.} \cite{li2002}). From
later observations ($>$ 1 day), the better limiting magnitude is
25.5 obtained by the Isaac Newton Telescope at La Palma (Salamanca
{\it et al.} \cite{salamanca2002}). The limiting magnitudes, summarized
in table 1, are displayed on figure \ref{curves}.

%%%%%%%%%%%%%%%%%%%%%%%%%%%%%%% ANALYSIS %%%%%%%%%%%%%%%%%%%%%%%%%
\section{Discussion}
\label{discussion}

Up to now, no afterglow of a short/hard GRB
was detected. However, it is
possible to get some constraints on the optical light curve. The
best limits to constrain the light curve for the afterglow of GRB
020531 comes from TAROT, KAIT, and INT data. If GRB 020531 was
followed by an optical afterglow, its light curve must lie in the
left part of figure \ref{curves}, below the dashed line.

Before GRB 020531, the earliest optical observations of a
short/hard GRB were obtained on GRB 010119 (Gorosabel {\it et al.}
\cite{Goro02}).

The decay slope index $\alpha$ for an afterglow of a short/hard
GRB (assuming flux
proportional to t$^{-\alpha}$) is now constrained
by GRB 020531 observations. Typical long GRBs afterglow decays are
between 0.7 and 1.8, marginally higher than 2
({\it i.e} GRB 980519, Vrba {\it et al.} \cite{Vrba00}).
Concerning GRB 020531, if
the flux of the afterglow was about the limiting magnitude of TAROT (R =
18.5 at 1.47 hour after the burst), then its decay slope $\alpha$
must be $>$ 2.2. If the afterglow was
fainter at this date, the decay slope should have a lower value.
%Anyway, a lower limit value is constrained by late observations
%to be $\alpha$ $>$ 0.4.
%Upper value of $\alpha$ is not constrained by
%the set of GRB 020531 observations.

Comparing to the dimest long GRBs, {\it e.g.} GRB 020124
(Berger {\it et al.} \cite{Berger02}, see figure \ref{curves}),
it implies that
the afterglow of GRB 020531 must be fainter.
TAROT upper limit is the first measurement obtained at such
early stage for a short/hard
GRB. It constrains the afterglow to be very dim.
This result is correlated to the 50-300 keV fluence which
is one decade fainter than typical those of long GRBs.

If the afterglow exists and decays with a t$^{-\alpha}$ law,
and if the source flux was about the limiting magnitude
of late observations, one can calculate R=22.0
at 1.47 hour after the GRB (TAROT observations) assuming
$\alpha$=1.2 (the typical case). Obviously, the afterglow can
be even fainter if it is dimer than the limiting magnitude
of late observations. As a consequence, plans for future searches
of afterglows of short/hard GRBs can be adressed: large aperture
telescopes, equiped by wide field cameras, should observe
early stages (until 1 hour after GRB).
Small aperture telescopes could also
contribute if they shoot until 15 min after GRB with a
limiting magnitude R $>$ 18.

\section{Conclusion}
\label{conclusion}

The afterglow of GRB 020531, if it exists, is very dim,
compared to the observed optical counterparts
of long GRBs.
If the optical counterpart of GRB 020531 is typical of short/hard GRBs,
it means that these kind of GRBs are associated
to very dim afterglows or no afterglow at all.
The observations suggest that the decay
slope $\alpha$ could be larger than 2.

It must be mentioned that dim afterglows can be
localized only by early optical observations (case of
GRB 020124 afterglow, found at 1.67 hour after the GRB).

Of course the possibility that GRB 020531 had no
afterglow cannot be excluded. This proves the need to get
more sensitive observations of the afterglow, as early as
possible after the main event. The TAROT observatory demonstrated
that this is possible, provided that the alert is sent quickly by
the instrument. The increase in the HETE performances, the recent
successful launch of the Curie-INTEGRAL satellite, as well as the
perspective of the SWIFT GRB dedicated satellite gives hope that
rapid observations of GRB optical counterparts will be soon
possible, as it was the case with BATSE (Akerlof {\it et al.}
\cite{Akerl99}, Bo¨\"er {\it et al.} \cite{Boer01}, Park {\it et al.}
\cite{Park1999}).

%If the source was of magnitude 20 at
%the same time, then its decay slope was larger than 1.6.
%This can be
%compared with the fastest decay indexes measured to date, e.g. 1.9
%for GRB 020124, which was considered as one of the faintest
%afterglow measured to date
%(R = 18.5, 22h after the GRB, Price et
%al. \cite{price2002})
%or 2.3 for GRB 980519 (Vrba {\it et al.}
%\cite{Vrba00}). In any case the afterglow of GRB 020531, if any
%should be either steeper, or far more fainter (by 5
%magnitudes at least) than any other afterglow observed to date.

\begin{acknowledgements}
The {\it T\'{e}lescope \`a Action Rapide pour les Objets
Transitoires} (TAROT) has been funded by the {\it Centre National
de la Recherche Scientifique} (CNRS), {\it Institut National des
Sciences de l'Univers} (INSU) and the Carlsberg Fundation. It has
been built with the support of the {\it Division Technique} of
INSU (INSU/DT). The TAROT CCD camera was built by a collaboration
between the CESR and the CEMES. We thank the technical staff
associated with the TAROT project: G. Bucholtz, J. Esseric, A.
Mayet, A.M. Moly, M. Nexon, H. Pinna, and C. Pollas.
\end{acknowledgements}
%%%%%%%%%%%%%%%%%%%%%%%%%%%%%%%%% REFERENCES %%%%%%%%%%%%%%%%%%%%%%%


\begin{thebibliography}{}

\bibitem[1999]{Akerl99} Akerlof, C., {\it et al.}, 1999, Nat 398, 400

\bibitem[1997]{Bar97}
Barthelmy, S., 1997,  Proceedings of the 4th Huntsville
Symposium, AIP conf. proc. 428, edts.\ C.A. Meegan, R.D. Preece,
and T.M. Koshut, p.\ 99.

\bibitem[2002]{Berger02}
Berger, E., {\it et al.}, 2002, ApJ to be published

%\bibitem[2000]{Berna00} Bernabei, S., {\it et al.}, 2000, GCNC 599

% \bibitem[1998a]{Bloom98a} Bloom, J.S., {\it et al.}, 1998a, GCNC 87

% \bibitem[1998b]{Bloom98b} Bloom, J.S., {\it et al.}, 1998b, ApJ 508, L21

\bibitem[1999]{Boer99}
Bo\"er, M., {\it et al.}, 1999, A\&AS 138, 579

\bibitem[2000]{Boer00} Bo\"er, M., and Gendre, B., 2000, A\&A 361,
L28

\bibitem[2001]{Boer01} Bo\"er, M., {\it et al.}, 2001, A\&A 378, 76

\bibitem[2002]{boer2002}
Bo\"er, M., {\it et al.}, 2002, GCNC 1408

\bibitem[2001]{bringer2001}
Bringer, M., Bo\"er, M., Peignot, C., Fontan, G., Merce, C., 2001,
Exper. Astrophys 12, 34

\bibitem[2002]{butler2002}
Butler, N. {\it et al.} 2002, GCNC 1415


% \bibitem[1999]{Castro99} Castro-Tirado, A.J., {\it et al.}, IAUC 7332

% \bibitem[1997]{Costa97} Costa, E., {\it et al.} 1997, Nat. 387, 783

% \bibitem[1999]{Costa99} Costa, E., 1999, A\&AS 138, 425

 \bibitem[1996]{dezalay96} Dezalay, J.P.., {\it et al.}, 1996, ApJ 471,
L27

% \bibitem[1998]{Diercks98} Diercks, A., {\it et al.}, 1998, ApJ 503, L105

% \bibitem[1997]{Djo97} Djogovski, S.G., {\it et al.}, 1998, IAUC 6660

% \bibitem[1998]{Djo98} Djogovski, S.G., {\it et al.}, 1998, GCNC 117

% \bibitem[1999]{Djo99} Djogovski, S.G., {\it et al.}, 1999, GCNC 510

% \bibitem[1999]{Dolan99} Dolan, C., {\it et al.}, 1999, GCNC 486

% \bibitem[1989]{Fish89} Fishman, G., {\it et al.}, 1989, Proc. Gamma-Ray
%Observatory science workshop, W.N. Johnson edt., GSFC Greenbelt,
%p.\ 2

%\bibitem[1997]{Fruch97} Fruchter, A., {\it et al.}, 1997, IAUC 6674

%\bibitem[2000]{Fyn00} Fynbo, J.P.U., {\it et al.}, 2000, GCNC 576

%\bibitem[1998]{Gal98} Gal, R.R., {\it et al.}, 1998, GCNC 88

%\bibitem[1997]{Gal97} Galama, T. {\it et al.}, 1997, IAUC 6655

%\bibitem[1999]{Gal99} Galama, T. {\it et al.}, 1999, Nat 398, 394

%\bibitem[2000]{Galy00} Gal-Yam, A., {\it et al.}, 2000, GCNC 593

%\bibitem[1997]{Gar97} Garcia, L., {\it et al.}, 1997, IAUC 6661

%\bibitem[1999]{Garna99} Garnavich, P., {\it et al.}, 1999, GCNC 495

%\bibitem[2000]{Garna00} Garnavich, P., {\it et al.}, 2000, GCNC 581

%\bibitem[2000]{Gehrels00}
%Gehrels, N., 2000,  Proceedings of the 5th Huntsville Symposium,
%AIP conf. proc. 526, edts.\ R.M. Kippen, S. Mallozzi, and G.
%Fishman, p.\ 671.

\bibitem[2002]{Goro02} Gorosabel, J., {\it et al.}, 2002, A\&A 383, 112

%\bibitem[1997]{Guar97} Guarnieri, A., {\it et al.}, 1997, A\&A 328, 13

%\bibitem[1998]{Halp98} Halpern, J.P., {\it et al.}, 1998, GCNC 134

%\bibitem[2000]{Halp00} Halpern, J.P., {\it et al.}, 2000, ApJ, {\it in press}

%\bibitem[1999]{Hari99} Harrison, F.A., {\it et al.}, 1999, ApJ 523, L121

%\bibitem[1999]{Henden99} Henden, A., {\it et al.}, 1999, GCNC 473

%\bibitem[2000]{Henden00} Henden, A., 2000, GCNC 652

%\bibitem[2000]{HendKlo00} Henden, A., and Klose, S., 2000, GCNC 656

\bibitem[2002]{henden2002}
Henden, A. {\it et al.} 2002, GCNC 1422

%\bibitem[1998]{Hjort98} Hjorth, J., {\it et al.}, 1998, GCNC 109

\bibitem[2002a]{hurley2002a}
Hurley, K. {\it et al.} 2002, ApJ 567, 447

\bibitem[2002b]{hurley2002b}
Hurley, K. {\it et al.} 2002, GCNC 1461

%\bibitem[1999a]{Jensen99a} Jensen, B.L., {\it et al.}, 1999, GCNC 454

%\bibitem[1999b]{Jensen99b}Jensen, B.L., {\it et al.}, 1999, GCNC 498

%\bibitem[1999]{Jha99}Jha, S., {\it et al.}, 1999, GCNC 476

\bibitem[2002]{klotz2002}
Klotz, A. {\it et al.} 2002, GCNC 1420

\bibitem[1993]{kou93} Kouveliotou, C., {\it et al.}, 1993, ApJ 403, L101

\bibitem[2000]{Kum00} Kumar, P., and Panaitescu, A., 2000, ApJ 541, L9

\bibitem[2000]{Kum2000b} Kumar, P., and Piran, T., 2000, ApJ 535,
    152

%\bibitem[2000]{kuu00} Kuulkers, E., {\it et al.}, P., 2000, ApJ 538, 638

\bibitem[2002]{lamb2002}
Lamb, D.Q. {\it et al.} 2002, in preparation

\bibitem[2002]{li2002}
Li, W. {\it et al.} 2002, GCNC 1405

%\bibitem[1999]{Mas99} Masetti, N., {\it et al.}, 1999, GCNC 462

\bibitem[1997]{Mesz97} M\'esz\'aros, P., and Rees, M., 1997,
      ApJ 476, 232

%\bibitem[1997]{Metz97} Metzger, M., 1997, IAUC 6588

%\bibitem[1997]{Mign97} Mignoli, M., 1997, IAUC 6661

%\bibitem[2000]{Mirabal00} Mirabal, N., 2000, GCNC 653

%\bibitem[2000]{Mohan00} Mohan, V., {\it et al.}, 2000, GCNC 595

%\bibitem[1999]{Pacie99} Paciesas, W.S., {\it et al.}, 1999, ApJS 122,

\bibitem[1998]{Pana98} Panaitescu, A., M\'esz\'aros, P., and Rees, M.,
   1998, ApJ 503, 314

\bibitem[1999]{Park1999} Park, H.S., {\it et al.}, 2001, A\&ASS 138, 577

%\bibitem[1998]{Peder98a} Pedersen, H. {\it et al.}, 1998, ApJ 503, 314

%\bibitem[1997]{Pedi97} Pedichini, F., {\it et al.}, 1997, A\&A 327, 36

%\bibitem[1999]{Pian99} Pian, E., {\it et al.}, 1999, A\&AS, 138, 463

\bibitem[1999]{Piran99} Piran, T., 1999, Phys. Rep. 314, 575

\bibitem[2001]{Piro2001} Piro, L., 2001, GCNC 932

%\bibitem[2001]{price2002} Price, P., 2002, GCNC 1221

\bibitem[1992]{Rees92} Rees, M., and M\'esz\'aros, P., 1992,
   MNRAS 258, 41

%\bibitem[1999]{Rei99} Reichart, D.E., {\it et al.}, 1999, ApJ 517, 692

\bibitem[2000]{ricker2000}
Ricker, G. R. \& HETE Science Team 2000, American Astronomical
Society Meeting 197, 2501

%\bibitem[2001]{Ricker01} Ricker, G.R., {\it et al.}, 2001, A\&AS {\it in
%press}

\bibitem[2002]{ricker2002}
Ricker, G. R. {\it et al.} 2002, GCNC 1399

%\bibitem[2000]{Sagar00} Sagar, R., {\it et al.}, 1999, Bull. Astron. Soc.
%India 28, 15

\bibitem[2002]{salamanca2002}
Salamanca, I. {\it et al.} 2002, GCNC 1433

\bibitem[1999]{Shanthi99}
shanthi, K. {\it et al.} 1999, Bull. Astr. Soc. India 27, 195

%\bibitem[1998]{Schl98} Schlegel, D.J., Finkbeiner, D.P., and
%Davis, M., 1998, ApJ 500, 525

%\bibitem[1999]{Sta99} Stanek, K.Z., {\it et al.}, 1999, ApJ 522, L39

\bibitem[1997]{Vanpara97} van Paradijs, J., {\it et al.}, 1997, Nat
386, 686

%\bibitem[2000]{paradijs2000}
%van Paradijs, J., Kouveliotou, C., Wijers, R. 2000, ARAA 38, 379

%\bibitem[2000]{Veillet00} Veillet, C., and Bo\"er, M., 2000, GCNC 598

\bibitem[2000]{Vrba00} Vrba, F., {\it et al.}, 2000, ApJ 528, 254

%\bibitem[2002]{wei2002}
%Wei, D.M., \& Lu, T. 2002 A\&A, 381, 731

\end{thebibliography}
\end{document}